\documentclass[a4paper,fleqn,usenatbib]{mnras}

\usepackage{newtxtext,newtxmath}

\usepackage[T1]{fontenc}
\usepackage{ae,aecompl}
\usepackage{hyperref}

\usepackage{graphicx}	
\usepackage{amsmath}	
\usepackage{amssymb}	

\title[Detecting the non-stationarities of GRB
jets]{A statistical method to detect non-stationarities of gamma-ray burst jets}

\author[Budai, Borgulya, Dawes, Szeifert, Varga and Raffai]{
A. Budai,$^{1}$\thanks{E-mail: arandras@caesar.elte.hu}
P. Raffai,$^{1, 2}$
B. Borgulya,$^{3}$
B. A. Dawes,$^{4}$
G. Szeifert,$^{1}$
and V. Varga$^{5}$
\\
$^{1}$E\"{o}tv\"{o}s University, Institute of Physics, 1117 Budapest, Hungary\\
$^{2}$MTA-ELTE Extragalactic Astrophysics Research Group, 1117 Budapest, Hungary\\
$^{3}$University of Edinburgh, School of Physics and Astronomy, EH9 3JZ, Edinburgh, UK\\
$^{4}$Columbia University, Department of Physics, New York, NY 10027, USA\\
$^{5}$Budapest University of Technology and Economics, Institute of Physics, 1111 Budapest, Hungary
}

\date{Accepted XXX. Received YYY; in original form ZZZ}

\pubyear{2019}
\begin{document}
\label{firstpage}
\pagerange{\pageref{firstpage}--\pageref{lastpage}}
\maketitle

\begin{abstract}
We propose a method to detect possible non-stationarities of gamma-ray burst jets. Assuming that the dominant source of variability in the prompt gamma light curve is the non-stationarity of the jet, we show that there should be a connection between the variability measure and the characteristic angle of the jet derived from the jet break time of the afterglow. We carried out Monte Carlo simulations of long gamma-ray burst observations assuming three radial luminosity density profiles for jets and randomizing all burst parameters, and created samples of gamma light curves by simulating jets undergoing Brownian motions with linear restoring forces. We were able to demonstrate that the connection between the variability and the characteristic angle is an anti-correlation in case of uniform and power-law jet profiles, and a correlation in case of a Gaussian profile.  We have found that as low as $50$ $(144)$ gamma-ray burst observations with jet angle measurements can be sufficient for a $3\sigma\ (5\sigma) $ detection of the connection. The number of observations required for the detection depends on the underlying jet beam profile, ranging from 50 (144) to 237 (659) for the four specific profile models we tested.
\end{abstract}




\section{Introduction}\label{sec:intro}

Gamma-ray bursts (GRBs) are among the most energetic events in the Universe. The equivalent isotropic energy emitted during a burst can reach up to $10^{52}$ ergs \citep{Kulkarni98, Meszaros2006}, and the typical radiated photon energy is in the  $100\mathrm{keV}-1\mathrm{MeV}$ range \citep{Klebesadel73, Zwart2001}. It is thought that GRBs are related to the death of massive stars and the coalescence of compact objects, where these two types of progenitors  produce two different classes of bursts distinguished mainly by their duration and spectral hardness: long GRBs and short GRBs, respectively \citep{Kouveliotou93, Kumar2015}. Long GRBs are often followed by long-lived emission in X-ray, optical and radio wavelengths. This so-called afterglow is rarely seen in the case of short GRBs, which is a further proof of their differing nature \citep{Kumar2015}. 

There are strong reasons to assume that GRB outflows are beamed in collimated jets \citep{Granot2007}, which reduces the high energy requirement compared to isotropic emission \citep{Rhoads97, Kumar2015}. The exact structure of jets is debated, usually axisymmetry is assumed, in which the luminosity density (i.e. luminosity per unit solid angle), $\epsilon$, varies with $\theta$, the angular distance measured from the jet axis \citep*{Zhang2002,Rossi2002,Granot2007}. For simplicity, a uniform jet profile is widely used in GRB studies, where $\epsilon$ is constant  within a well-defined $\theta_\mathrm{j}$ half opening angle (which changes from burst to burst), and zero otherwise \citep{Rossi2002,Granot2007}. Rossi et al., however, proposed a structured jet profile, where $\epsilon$ drops off gradually with $\theta$ \citep{Rossi2002}. Here, the exact beam profile can be different in different models: $\epsilon$ as a function of $\theta$ can follow a power-law \citep{Rossi2002} or a Gaussian curve \citep{Zhang2002}. The power-law profile is usually assumed to be exactly the same for all jets (\textit{universal structured jet}), while the Gaussian profile is expected to vary for different GRBs (\textit{quasi-universal structured jet}) \citep{Kumar2015}.

One phenomenon interpreted as an evidence for the existence of jets is the achromatic break (the so-called \textit{jet break}) seen in many afterglow light curves \citep{Kumar2015}.  In the uniform jet profile model, the time elapsed between the detection of the GRB and the break (the so-called \textit{jet break time}, $t_\mathrm{b}$) depends on $\theta_\mathrm{j}$, while in the structured jet model it depends on the $\theta_\mathrm{v}$ viewing angle between the line-of-sight and the jet axis \citep{Rossi2002,Zhang2002}. Thus, one can deduce the same characteristic angle, $\theta_\mathrm{c}$, of the GRB from $t_\mathrm{b}$ \citep[for details see e.g.][]{Wang18}, and interpret it as either $\theta_\mathrm{c} \equiv \theta_\mathrm{j}$ or $\theta_\mathrm{c} \equiv \theta_\mathrm{v}$, depending on the model used \citep{Kumar2015}. 

Light curves of GRBs -- the gamma photon counts as a function of time -- have high variabilities \citep{Strong74, Kumar2015}. The usual explanation for this is given by the internal shock model, in which the radiation is emitted by pairs of shells colliding with each other creating single pulses in each of these collisions \citep{Rees94, Kumar2015}. An alternative  explanation for the peaks in the light curves assumes the non-stationarity (i.e. movement) of jets \citep*{Roland1994}. For example, Portegies Zwart et al. tried to explain the complex light curves of GRBs with the precession of jets \citep*{Zwart1999}. In their model, the precession of a tilted accretion disk around a black hole results in the precession of the jet, which causes peaks to appear in the gamma light curve every time the jet crosses the line-of-sight. In this way, multiple crosses can create a complex light curve. Since this process is deterministic, the temporal structure of light curves can be used to infer parameters of the movement by fitting them with the model.  Other studies tried to elaborate on this in order to simulate more realistic light curves \citep{Zwart2001,Lei2007}. Since these models always involve several free parameters, they provide possible explanations for the variabilities of light curves, but they cannot determine whether jets are stationary or not in reality, even if the observed prompt light curves can be reproduced with them. Nevertheless, as an independent confirmation, Liska et al. have recently demonstrated using 3D magnetohydrodynamic  simulations that the plausible configuration of a tilted disk around a black hole can indeed result in the precession of jets \citep{Liska2017}.

In this paper we introduce a statistical method that can be used to detect non-stationarities of jets by searching for a possible connection between two measurable quantities in a sample of GRBs, i.e. the variabilities of prompt gamma light curves and the characteristic angles derived from jet break times of GRB afterglows. We show that if the dominant source of variability in the prompt gamma light curve is the non-stationarity of the jet, the connection should exist between the two quantities in the form of a correlation or anti-correlation depending on the jet model we assume to be realistic. To test this idea, we used Monte Carlo simulations of various numbers of GRB observations, measured the variabilities of the simulated light curves produced by non-stationary jets, and correlated them with the corresponding $\theta_\mathrm{j}$ or $\theta_\mathrm{v}$ values. Using this method, we also measured the lowest number of GRBs with afterglows that need to be observed in order to detect the connection with a $3\sigma$ significance.

Here we would like to emphasise that we do not want to argue in favour of the non-stationarity of jets, nor against it - we only propose a method to test it - and thus there are some theoretical issues that our paper does not cover: for example whether this model can explain what causes the coherent episodes called pulses, why the first pulse of light curves is often the most luminous, or what causes the softening of light curves. We would like to encourage the reader to check the papers by the authors arguing in favour of the non-stationary jet model. Some of their works were already mentioned above.  

There are two major caveats concerning our work that have to be taken into account. First, it is possible that a movement of the GRB jet can affect the light curve of the afterglow, which makes the interpretation of the jet break (and thus the measurement of the characteristic angle) less certain. By assuming that the movement of the jet is perturbative, we neglect this potential effect here, and leave it for future investigations. Also, a moving jet, in principle, generates a $\theta_\mathrm{v}$ that is changing with time, which questions the original definition of $\theta_\mathrm{v}$ that assumes a fixed viewing angle for the jet. In order to overcome this technical issue, we assume $\theta_\mathrm{v}$ to be the initial viewing angle of the non-stationary jet.

Our paper is organised as follows. In Section \ref{sec:explmodel}, we describe how non-stationarities of jets can create a connection between characteristic angles and variabilities of light curves of GRBs. We introduce our method for measuring the variability of light curves in Section \ref{sec:var}. We present the way we randomised GRB parameters in our Monte Carlo simulation in Section \ref{sec:param}. In Section \ref{sec:result}, we present our simulation results, and in Section \ref{sec:conc}, we offer our conclusions.

\section{Potential connection between light curve variabilities and characteristic angles}\label{sec:explmodel}

The non-stationarity of jets can, in principle, explain the complex temporal structure of GRB light curves: as the jet sweeps through space, the change in the viewing angle can cause a change in the observed gamma flux as well, resulting in a variable light curve \citep[see e.g.][]{Zwart1999}. At the same time, however, the $\epsilon(\theta)$ luminosity density of a jet inevitably has an intrinsic time dependence at all $\theta_\mathrm{v}$ viewing angles, determined by the time dependence of the emission process. This means, that the variability of an observed light curve is always the combined result of the time-dependent emission and the possible movement of the jet.

It is reasonable to assume that modelling $\epsilon(\theta)$ as $\epsilon(\theta)=const.$ for all $\theta \leq \theta_\mathrm{j}$ (i.e. the uniform jet profile model) is an oversimplification when considering the angular dependence of the luminosity density function, and that $\epsilon(\theta)$ can more realistically be described by other, non-trivial and continuous functions of $\theta$. For such models of $\epsilon(\theta)$, the $\theta_\mathrm{c}$ characteristic angle that we can derive from the jet break time is typically the viewing angle, i.e. $\theta_\mathrm{c} \equiv \theta_\mathrm{v}$ \citep[see e.g.][]{Zhang2002}. If the movement of the jet is indeed perturbative, then the light curve variability caused solely by this motion is determined by the steepness of the $\epsilon(\theta)$ curve at $\theta_\mathrm{v}$: the steeper the curve is, the larger the light curve variability is. For jet profile models where the steepness of the $\epsilon(\theta)$ curve is a monotonous function of $\theta_\mathrm{v}$ (which is the case for e.g. the power-law profile, and for the Gaussian profile for either all $\theta_\mathrm{v} \geq \sigma$ or all $\theta_\mathrm{v} < \sigma$), the variability should be connected to $\theta_\mathrm{v}$. The connection being a correlation or an anti-correlation depends on whether the steepness of $\epsilon(\theta)$ increases or decreases, respectively, with $\theta$. Note that for a set of real observations, we must define a measure of the gamma light curve variability that is normalized for the net effect of the different distances and total energy outputs of individual GRBs, both of which are unknown from the gamma observations. Introducing a normalization factor in the variability measure $\mathcal{V}$ that carries out this compensation can change the nature and strength of the $\mathcal{V}-\theta_\mathrm{v}$ connection, nevertheless it should not change the fact that the connection exists.

In the less realistic case when the correct model for $\epsilon(\theta)$ is the uniform jet profile, the characteristic angle is either $\theta_\mathrm{c} \equiv \theta_\mathrm{j}$ \cite[see e.g.][]{Rossi2002} or $\theta_\mathrm{c}\equiv \theta_\mathrm{j} - \theta_\mathrm{v}$ \cite[see e.g.][]{Wei2003}. In this case, a moving jet can point repeatedly towards and away from the observer (the latter meaning that $\theta_\mathrm{v}$ becomes larger than $\theta_\mathrm{j}$ multiple times), which results in multiple peaks appearing in the observed light curve. Under these circumstances, jets with larger $\theta_\mathrm{j}$ allow observers to spend more time inside jets. This results in an anti-correlation between light curve variabilities and $\theta_\mathrm{c}$ values, regardless of which interpretation we accept for $\theta_\mathrm{c}$ from the above two cases. 

As we will show in this paper, assuming that on a given time scale (e.g. on the time scale much shorter than the total duration of the GRB) the variability is dominated by the movement of the jet, there should be a connection (a correlation or anti-correlation, depending on the jet profile) between the variability of the light curve measured on this time scale, and the characteristic angle of the jet derived from the afterglow. Note that without a moving jet, when only the time-dependent emission process is considered, no such connection between the gamma variability and the characteristic angle is expected or proposed.

\section{Measuring gamma light curve variabilities} \label{sec:var}

There is a well known standard method for deriving the characteristic angles of GRB jets, $\theta_\mathrm{c}$, from jet break times of GRB afterglows
\citep[see e.g.][]{Wang18}. However, for measuring the variability of a gamma light curve, no such standard method exists. For example, Fenimore et al. used a boxcar filter to smooth the light curve, and took the average difference between the original and the smoothed curve as a variability measure \citep{Fenimore2000}. Li et al. used a Stavitzky-Golay filter for the same purpose \citep{Li2006}. Note that both groups of authors assumed that GRB jets are stationary, and that variabilities of GRB light curves are the result of the intrinsic time dependence of the gamma emission process.

Following the consensus in papers dealing with non-stationary jets \citep[see e.g.][]{Zwart1999, Lei2007}, we also assumed that GRBs with stationary jets do not have multiple local maxima and minima in their detected light curves, but instead, their light curves rise from a known background level to a global maximum, and fade back to the background level again. We considered all other types of variabilities of light curves (i.e. multiple local maxima and minima) to be dominated by the movements of jets. With this, we defined our variability measure, $\mathcal{V}$, in a way to satisfy the following two requirements: (i) $\mathcal{V}$ should have a value of zero when the light curve does not have multiple local maxima and minima, and (ii) the exact form of the intrinsic time dependence of $\epsilon(\theta)$ should have a negligible effect on $\mathcal{V}$. Our method for measuring light curve variabilities satisfies both conditions.

\begin{figure}
	\includegraphics[width=\columnwidth]{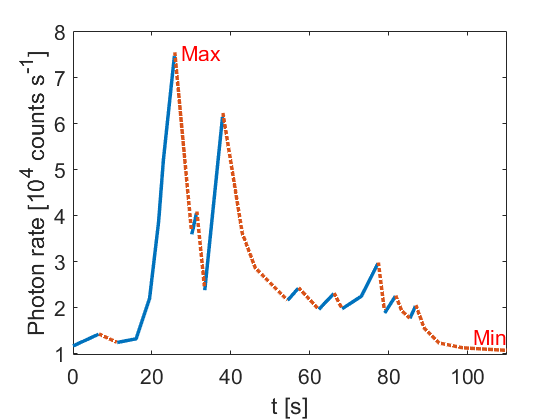}
    \caption{
    	An example gamma light curve, corresponding to GRB990123 or BATSE Trigger 7343 (see \citealt{BATSE} for details). The red dotted segments denote the descending parts of the curve adding positive contribution to our $\mathcal{V}$ variability metric (see Eq.(\ref{eq:var})). Notice that the sum of the heights of the red dotted segments is larger than the difference between the maximum (labelled as \emph{Max}) and  minimum (labelled as \emph{Min}) value of the curve.
    }
    \label{fig:example_peaks}
\end{figure}

We treat the background-subtracted GRB light curves as time series of $F_{i}$ photon fluxes measured in $N_\mathrm{bin}$ number of successive time bins, where $N_\mathrm{bin}$ depends on the ratio of the total duration of the GRB and the time resolution of the detector. With these, we defined $\mathcal{V}$ as:
\begin{equation}
\mathcal{V} = -\frac{1}{T_{90} F_\mathrm{max}} \left[ F_\mathrm{max}-\sum_{i=1}^{N_\mathrm{bin}-1} \left(F_{i}-F_{i+1} \right) \times H\! \left( F_{i}-F_{i+1} \right)  \right]
\label{eq:var}
\end{equation}
where $F_i$ is the photon flux measured in the $i$th time bin, $F_\mathrm{max}$ is the maximum value of the light curve, $H(\cdot)$ is the Heaviside step function, and $T_\mathrm{90}$ is the time interval between the epochs when 5 per cent and 95 per cent of the total fluence is registered by the detector \cite[see e.g.][]{Kumar2015}.

Fig. \ref{fig:example_peaks} shows a real example for a GRB light curve (GRB990123 or BATSE Trigger 7343; see \citealt{BATSE} for details), where we highlighted the descending segments of the curve with a red dotted curve style. Note that according to Eq.(\ref{eq:var}), only the heights of these descending segments contribute to the summation given in the definition of $\mathcal{V}$. Eq.(\ref{eq:var}) also shows that the duration of the GRB is compensated through normalising the metric by the measured $T_\mathrm{90}$ of the GRB, and thus, $\mathcal{V}$ depends on the ascending parts and the exact shape of the light curve only through $T_\mathrm{90}$.

It is worth mentioning here, that in our simulations of GRB observations, we neglected the effect the curvature of a non-stationary jet front can potentially have on the shape of peaks in the light curve. Portegies Zwart et al. suggested that this effect can result in individual peaks having fast rise and shallow decay pulse profiles \cite[see e.g.][]{Zwart2001}. Note that this is a secondary effect, and the way our variability metric is defined in Eq.(\ref{eq:var}) makes $\mathcal{V}$ robust to the exact shapes of peaks in GRB light curves.

\section{Parameters of gamma-ray bursts}\label{sec:param}

In this section we introduce the relevant parameters of GRBs we implemented to our simulations, and the distributions we associated to them. In subsection \ref{subsec:ProfsAngles}, we describe the three different luminosity density profiles, and the corresponding distributions of angular parameters we applied. In subsection \ref{subsec:time}, we discuss what intrinsic time dependence of GRB emissions we assumed. Subsection \ref{subsec:Z} considers the redshift distribution of long GRBs.

\subsection{Jet profiles and angles}\label{subsec:ProfsAngles}

We used three different jet profiles in our tests: the uniform, the power-law, and the Gaussian profile (see Section \ref{sec:intro}). 
In the two uniform jet profile models (one assuming $\theta_\mathrm{c}=\theta_\mathrm{j}$, and the other assuming $\theta_\mathrm{c}=\theta_\mathrm{j}-\theta_\mathrm{v}$), the luminosity density $\epsilon_\mathrm{u}$ as a function of the angle measured from the centre of the jet, $\theta$, is defined as: 
\begin{equation}
\epsilon_\mathrm{u}(\theta) = \begin{cases}
\epsilon_\mathrm{0,u} & \mathrm{for}\ \theta \in[0, \theta_\mathrm{j}] \\
 0 & \mathrm{otherwise},
\end{cases}
\label{eq:epsHom}
\end{equation} 
where $\epsilon_\mathrm{0,u}$ is a constant, and $\theta_\mathrm{j}$ is the half opening angle of the jet. According to e.g. \cite{Ghirlanda2013}, $\theta_\mathrm{j}$ values follow a log-normal distribution in nature, with a probability density function (PDF) of:
\begin{equation}
\rho_\mathrm{j}(\theta_\mathrm{j}) = \frac{1}{\sqrt{2\pi} \sigma_\mathrm{\theta}  \theta_\mathrm{j}} \exp\left( -\frac{(\ln \theta_\mathrm{j}-\mu_\mathrm{\theta})^2}{2\sigma_\mathrm{\theta}^2}\right),
\end{equation}
where $\mu_\mathrm{\theta}=1.742$, $\sigma_\mathrm{\theta} = 0.916$  \citep{Ghirlanda2013}. 
We also let the initial viewing angle, $\theta_\mathrm{v}$ change from burst to burst, but in the case of the uniform profile, we demanded $\theta_\mathrm{v}$ to be $\theta_\mathrm{v}<\theta_\mathrm{j}$ in order for the GRB to be detected. The corresponding PDF for $\theta_\mathrm{v}$ values is:
\begin{equation}
\rho_\mathrm{v}(\theta_\mathrm{v}) = \frac{\sin(\theta_\mathrm{v})}{1-\cos(\theta_\mathrm{j})}
\end{equation}
for $\theta_\mathrm{v} \in [0,\theta_\mathrm{j}]$, and $\rho_\mathrm{v}=0$ otherwise.

The power-law profile model suggests that GRB jets have the following luminosity density profile:
\begin{equation}
\epsilon_\mathrm{p} (\theta) = \begin{cases}
\epsilon_\mathrm{0,p} & \mathrm{for}\ \theta \in[0, \theta_\mathrm{0}] \\
\epsilon_\mathrm{0,p} \left(\frac{\theta}{\theta_\mathrm{0}}\right)^{-2} & \mathrm{otherwise},
\end{cases}
\label{eq:epsPow}
\end{equation}
where $\epsilon_\mathrm{0,p}$ is a constant characterizing the luminosity density at and near the centre of the jet, and $\theta_\mathrm{0}$ is a small angle introduced to avoid singularity near the jet axis \citep{Rossi2002}. In our simulations, we let $\theta_\mathrm{0}$ randomly vary from burst to burst, with a uniform distribution within the range $\theta_\mathrm{0} \in [0.5^\circ,0.8^\circ]$. We chose a cutoff viewing angle of $30^\circ$, above which we assumed that GRBs with power-law jet profiles remain undetected. This was motivated by the fact that the $\theta_\mathrm{c}$ derived from jet break time is $\theta_\mathrm{c} \equiv \theta_\mathrm{v}$ in the power-law profile model, and based on empirical data, $>95$ per cent of $\theta_\mathrm{c}$ are $\theta_\mathrm{c}<30^\circ$ \citep [see e.g.][]{Ghirlanda2013}. With this, the PDF of viewing angles for the power-law profile model is:
\begin{equation}
\rho_\mathrm{v}(\theta_\mathrm{v}) =\frac{\sin(\theta_\mathrm{v})}{1-\cos(30^\circ)},
\end{equation}
where  $\theta_\mathrm{v}\in [0^\circ, 30^\circ]$.

The third jet model we implemented is the quasi-universal structured jet with a Gaussian profile \citep{Zhang2002, Granot2007}. . For this, the luminosity density function is:
\begin{equation}
\epsilon_g(\theta) = \epsilon_\mathrm{0,g} \cdot \exp \left(-\frac{\theta^2}{2 \theta_\mathrm{0}^2}\right),
\label{eq:epsGauss}
\end{equation}
where $\theta_\mathrm{0}$ is a parameter of the GRB we chose from a uniform distribution in the range $\theta_\mathrm{0} \in [3^\circ, 5^\circ]$  \citep[see][]{Salafia2015}, and $\epsilon_\mathrm{0,g}$ is a constant characterising the luminosity density at the centre of the jet. 

\subsection{The time dependence of $\epsilon_\mathrm{0}$} \label{subsec:time}

As can be seen in Eq.\eqref{eq:epsHom}, \eqref{eq:epsPow} and \eqref{eq:epsGauss}, $\epsilon(\theta)$ has a universal form:
\begin{equation}
\epsilon(\theta) = \epsilon_\mathrm{0} f(\theta),
\label{eq:univEps}
\end{equation} 
where $f(\theta)$ is a given function of $\theta$, and the intrinsic time dependence of the GRB emission process is described by $\epsilon_\mathrm{0}$ being a function of time: $\epsilon_\mathrm{0}=\epsilon_\mathrm{0}(t)$.

In the case of a moving jet, the time dependence of $\epsilon$ cannot be observed directly; for example Portegies Zwart et al. used a light curve model that assumes an exponential rise, followed by a plateau and a stiff decay \citep{Zwart1999}. We used a Tukey window to model $\epsilon_\mathrm{0}(t)$:
\begin{equation}
\epsilon_\mathrm{0}(t) = \begin{cases}
\frac{\epsilon_\mathrm{max}}{2}\left \{1 + \cos\left(\frac{2 \pi}{q T}\left[t\! -\! \frac{q T}{2}\right]\right)\right \} &\!\!\! \mathrm{if}\ 0 \leq t < \frac{q T}{2} \\
\epsilon_\mathrm{max} &\!\!\! \mathrm{if}\ \frac{q T}{2} \leq t\! <\! 1\! -\! \frac{q T}{2} \\
\frac{\epsilon_\mathrm{max}}{2}\left \{1 + \cos\left(\frac{2 \pi}{q T}\left[t\! -\! 1\! +\! \frac{q T}{2}\right]\right)\right \} &\!\!\!
 \mathrm{if}\ 1\! -\! \frac{q T}{2} \leq t < T,
\end{cases}
\label{eq:tukey}
\end{equation}
where $\epsilon_\mathrm{max}$ is the maximum value of $\epsilon_\mathrm{0}$, $q$ is a number we chose from a uniform distribution in the range $q\in[0.3,0.47]$, and $T$ is the duration of the GRB in the comoving frame, which we simply chose from the log-normal distribution given for $T_\mathrm{90}$ values in \cite{Tarnopolski2016}:
\begin{equation}
\rho(T) = \frac{1}{\sqrt{2\pi}\sigma_\mathrm{T} T} \exp\left( -\frac{(\ln T-\mu_\mathrm{T})^2}{2\sigma_\mathrm{T}^2}\right),
\label{eq:T}
\end{equation}
where $\mu_\mathrm{T}=1.487$ and $\sigma_\mathrm{T}=0.326$. 

The way we defined $\mathcal{V}$ in Section \ref{sec:var} makes $\mathcal{V}$ completely independent from the exact shape of the curve for $\epsilon_\mathrm{0}(t)$ as long as the curve does not have multiple local minima and maxima. Thus the results we obtained with the Tukey window curve defined in Eq.(\ref{eq:tukey}) can be generalized to all $\epsilon_\mathrm{0}(t)$ functions satisfying this condition. Additionally, the fact that $\mathcal{V}$ is normalized with $F_\mathrm{max}$ in Eq.(\ref{eq:var}) makes $\mathcal{V}$ independent from $\epsilon_\mathrm{max}$, and thus we could set $\epsilon_\mathrm{max}=1$ for all our simulated GRB observations.

\subsection{Redshifts}\label{subsec:Z}

We selected the comoving distances, $d$, of our simulated GRBs from the following PDF (which describes a uniform distribution in volume within a sphere with radius $d_\mathrm{max}$):
\begin{equation}
\rho(d)=\frac{3 d^2}{4\pi d_\mathrm{max}^3},
\end{equation}
where $d_\mathrm{max}$ is the highest $d$, which we chose to be $d_\mathrm{max}=9.2\mathrm{Gpc}$, corresponding to a GRB detected at redshift $z=8.2$ \citep[see e.g.][]{Tanvir2009}. After randomizing $d$ for all simulated GRBs, we calculated the corresponding $z$ redshifts assuming a standard $\Lambda$CDM cosmology with density parameters $\Omega_\mathrm{\Lambda} = 0.714$, $\Omega_\mathrm{m} = 0.286$, and a Hubble constant of $H_0 = 69.6\ \mathrm{km\, s^{-1} Mpc^{-1}}$ \citep{Bennett14}. 

\section{The simulation and results} \label{sec:result}

In this section we describe the way we simulated the movement of jets (see Section \ref{subsec:movingjet}), and we discuss the results we obtained from our simulation (see Section \ref{subseq:results}).

\subsection{The movement of jets}\label{subsec:movingjet}

As we discussed in Section \ref{sec:explmodel}, the $\mathcal{V}-\theta_\mathrm{c}$ connection (to first order) depends predominantly on the slope of the $\epsilon(\theta)$ function if the movement of the jet is perturbative. This means that the fact that a connection (a correlation or anti-correlation) exists between $\mathcal{V}$ and $\theta_\mathrm{c}$ is (to first order) independent from the exact shape of the trajectory of the jet. Taking this into account, and the fact that the exact mechanism of the jet launching and evolution is not yet fully understood (e.g. the role of the magnetic field and the cocoon, or the effect of turbulence during the propagation of the ejected material through the envelope of the progenitor star; see e.g. \citealt{Kumar2015}), we used a simple toy model to describe motions of jets: we assumed a Brownian random angular motion for the jet with a linear restoring force, a model which ensures that the angular displacement of the jet remains small around $\theta_\mathrm{v}$. Note that this model is not based on any physical theory, but it indeed results with a random motion where $\mathcal{V}-\theta_\mathrm{c}$ connection only depends on the local steepness of the $\epsilon(\theta)$ function.

Instead of using the observer's frame, we fixed the GRB jet, and simulated the trajectory of the observer within the frame that has the centre of the jet as its origin. In this frame, the observer's motion is described by the Langevin-equation:
\begin{equation}
\frac{\mathrm{d^2}\mathbf{r}}{\mathrm{d}t^2}=\mathbf{F} - k\mathbf{r},
\label{eq:motion}
\end{equation}
where $\mathbf{r}$ is the position vector of the observer within the two dimensional plane of the jet's cross section, $\mathbf{F}$ is a two-component vector with components drawn from a normal distribution with zero mean and a standard deviation of 10 at every time step of the evolution, and $k$ is a spring constant randomly chosen for every GRB from a uniform distribution in the range $k\in [0.1, 25.1]$. Note that the components of all vectors are measured in angular units (in degrees), and that Eq.(\ref{eq:motion}) describes a motion with displacements around an average and initial $\theta_\mathrm{v}$ viewing angle. 

\begin{figure*}
	\includegraphics[width=\textwidth]{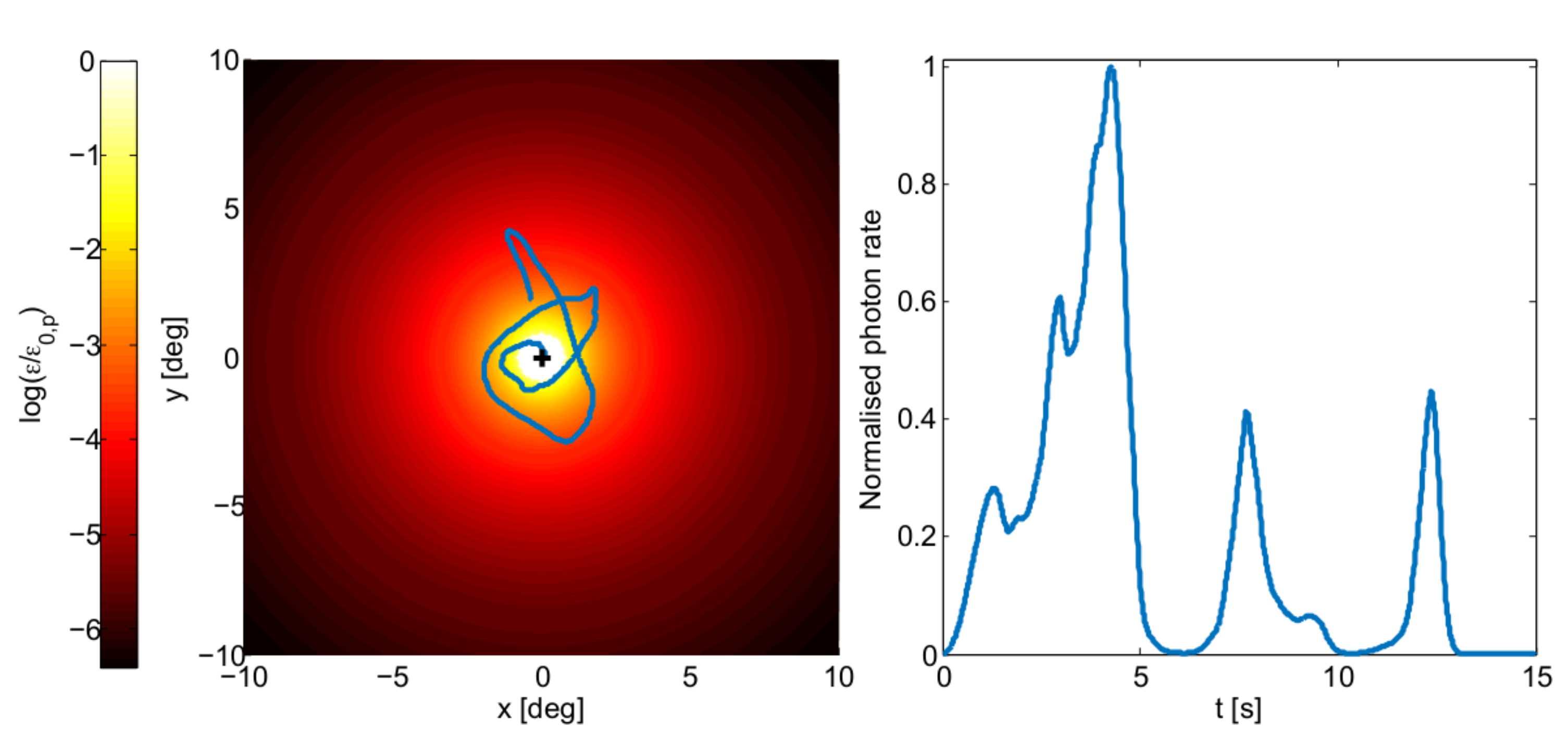}
	\caption{
	        This figure illustrates how a gamma light curve of a GRB with non-stationary jet is simulated in our analysis (for details of the simulation, see Section \ref{subsec:movingjet}). The colour map in the left panel shows the cross-section of a GRB jet with a power-law profile (see Eq.(\ref{eq:epsPow}) and Section \ref{subsec:ProfsAngles} for details). The simulated trajectory of the observer within the jet's reference frame (described in Section \ref{subsec:movingjet}) is shown with a blue line, and the starting point (i.e. the position of the observer at t = 0 seconds) is marked with a black cross near the centre of the jet. In the right panel we show the simulated light curve that the observer measures along the trajectory (with the intrinsic time-dependence of the emission taken into account; see Section \ref{subsec:time}), normalised by the highest photon rate value of the light curve. Note that in this illustration we disregarded the effect of the cosmological red shift.
	}
	\label{fig:jetCurve}
\end{figure*}

The simulated trajectory started at a $\theta_\mathrm{v}$ randomized for the GRB (see Section \ref{subsec:ProfsAngles} for details) and with zero initial velocity. We solved Eq.(\ref{eq:motion}) with the Euler-method, where we applied a constant time step of $\Delta t=0.01$ sec, and sampled $\epsilon(\theta)$ at each time step and along the trajectory, up until a duration of $T$ (see Section \ref{subsec:time}). The sampling resulted in a series of $F_i$ values (i.e. a light curve, see Section \ref{sec:var} for details and Figure \ref{fig:jetCurve} for an example), that we could use to calculate the $\mathcal{V}$ using the method described in Section \ref{sec:var}.

\subsection{Results}\label{subseq:results}

Using the method we described in Section \ref{subsec:movingjet}, we simulated 1000 independent samples of $N$ GRBs, where we carried out our analysis for every integer value of $N$ in the range of $N\in [50,1000]$. For each set of $N$ GRBs, we calculated the Pearson correlation coefficient, $r$, and the corresponding $p$-value, and constructed the distributions of $r$ and $p$ values from the 1000 independent samples. For each of the four different jet profile models, and for all $N$ values, we show the medians and the 68 per cent confidence limits of the $r$ and $p$ distributions in Figure \ref{fig:hlhr}. As references, we also indicate the $p_{3\sigma}=0.003$ and $p_{5\sigma}=6\times10^{-7}$ levels in Figure \ref{fig:hlhr}, and we give the $N$ values ($N_{3\sigma}$ and $N_{5\sigma}$) in Table \ref{tab:table1} for which the medians and the 97.8th percentiles of the $p$ distributions are equal to $p_{3\sigma}$ and $p_{5\sigma}$. Table \ref{tab:table1} also shows the $r_{\inf}$ values to which the medians of the $r$ distributions converge as $N$ increases. Note that $r_{\inf}$ (and $r$ values in general) are negative for the power-law and uniform profiles, indicating an anti-correlation between $\mathcal{V}$ and $\theta_\mathrm{c}$, while $r_{\inf}$ and $r$ values are positive for the Gaussian profile, indicating a correlation between $\mathcal{V}$ and $\theta_\mathrm{c}$. 

As discussed in Section \ref{sec:explmodel}, these results can be explained by the difference in how the steepness of the $\epsilon(\theta)$ function changes with $\theta$ within the different jet profiles. In case of a structured profile, the change of this steepness is what decides the type of the connection between the viewing angle and the variability of the light curve, provided that the movement is perturbative. If the steepness decreases with $\theta$, the connection will be an anti-correlation as in the case of the inverse power-law profile, and if it increases, it will be a correlation as in the case of the Gaussian profile if ($\theta_\mathrm{v} \leq \sigma$ for most GRB observations, which can reasonably be assumed due to selection effects) in most of the cases. In case of the uniform jet models, the connection is always an anti-correlation. A theoretical jet model, where no connection is expected in the framework of our theory, would be a linear structured jet model, i.e. where the steepness of $\epsilon(\theta)$ is constant with $\theta$. 
  
Note that, according to the results given in Table \ref{tab:table1}, as low as $N=50$ (144) GRB observations with both gamma light curves and derived $\theta_\mathrm{c}$ values can potentially be enough to detect a $\mathcal{V}-\theta_\mathrm{c}$ connection as a sign of non-stationarity of GRB jets with a $3\sigma\ (5\sigma)$ significance.

\section{Conclusions} \label{sec:conc}

In this article we proposed a method to detect possible non-stationarities of GRB jets. Assuming that the dominant source of variability in the prompt gamma light curve is the non-stationarity of the jet, we showed that there should be a connection between the variability measure, $\mathcal{V}$ (that we defined in Section \ref{sec:var}) and the characteristic angle of the jet, $\theta_\mathrm{c}$, derived from the jet break time of the afterglow (see Section \ref{sec:explmodel} for details).

We carried out Monte Carlo simulations of long GRB observations assuming three radial luminosity density profiles for jets and randomizing all burst parameters, and created samples of gamma light curves by simulating jets undergoing Brownian motions with linear restoring forces. We were able to demonstrate that the connection between the variability and the characteristic angle is an anti-correlation in case of uniform and power-law jet profiles, and a correlation in case of a Gaussian profile.

Besides the caveats we already discussed (e.g. the jet movements effecting the afterglow light curve and the value and interpretation of $\theta_\mathrm{c}$), three additional issues also need attention in the future, especially in real data analyses. First, we treated our virtual gamma detector as being equally sensitive to every GRB we simulated, and we also oversimplified the treatment of noise in detector data and the measurement errors affecting $\theta_\mathrm{c}$. We propose to carry out a more sophisticated analysis on these when the detectors and datasets to be used in real data processing become set in the future. Second, we demonstrated the applicability of our definition of $\mathcal{V}$ (see Section \ref{sec:var}) in detecting non-stationarities of GRB jets through testing for a potential $\mathcal{V}-\theta_\mathrm{c}$ connection, however we did not {\it optimize} our $\mathcal{V}$ for the detection of the connection in real life circumstances. In fact, carrying out such an optimization of the $\mathcal{V}$ metric could decrease the $N$ numbers given in Table \ref{tab:table1}, i.e. the minimum number of GRBs required for a high-confidence detection of jet non-stationarities. Finally, although our method for detection is robust to the different angular motions of GRB jets (see Section \ref{subsec:movingjet} for a detailed discussion), elaborating on the models of jet motions could lead to more realistic modelling of the potential $\mathcal{V}-\theta_\mathrm{c}$ connection. Note that all three of these issues are beyond the scope of this paper, and thus we leave all of them for future investigations.

\begin{figure*}
	\includegraphics[width=\textwidth]{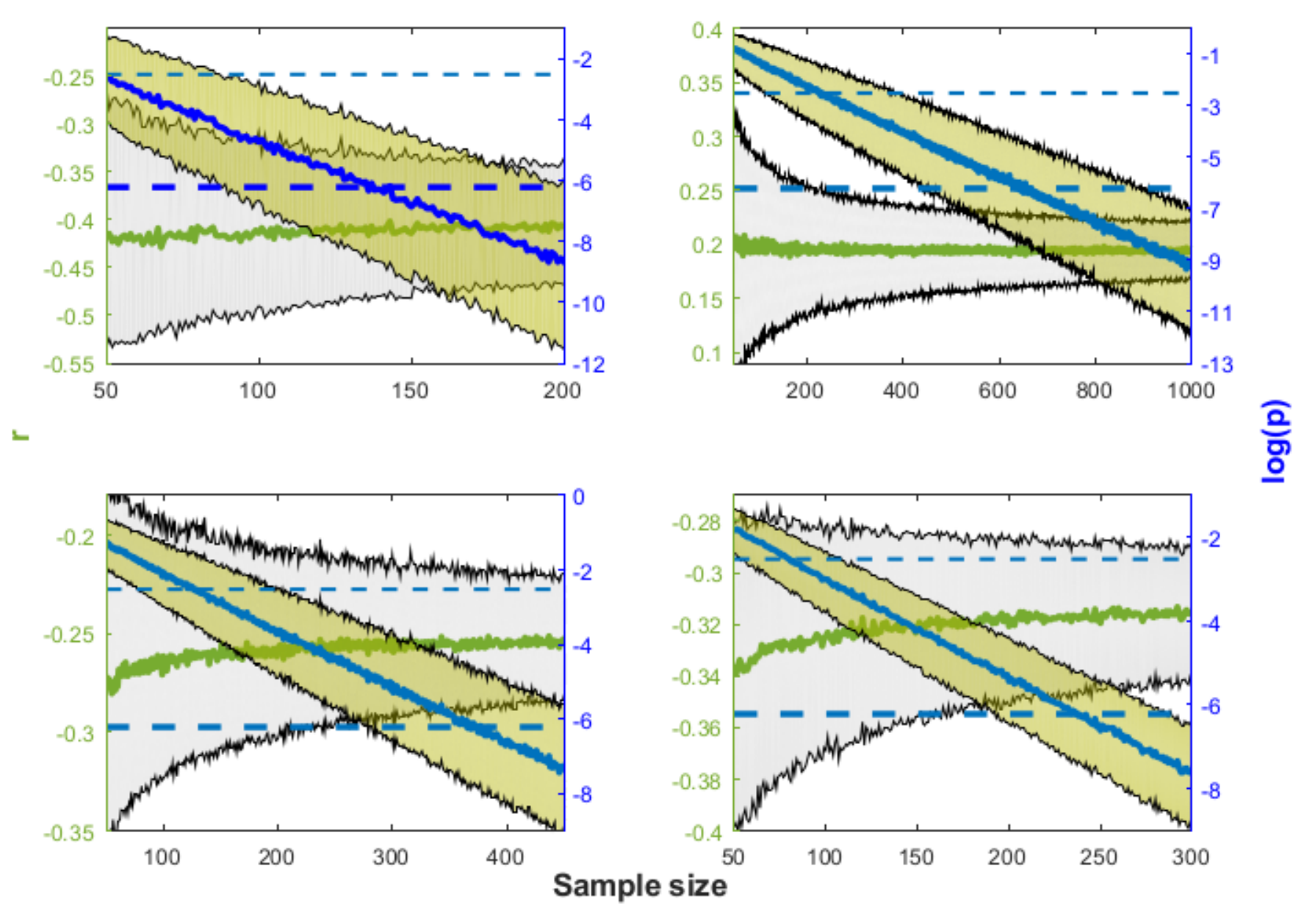}
	\caption{
		The figure shows, as a function of the sample size $N$, the medians (green and blue solid lines) and 68 per cent confidence intervals (shaded areas bound by black solid lines) of $r$ and $p$ distributions constructed from 1000 samples of $N$ simulated GRBs. We obtained the $r$ Pearson correlation coefficients and the corresponding $p$ values by cross-correlating the light curve variability measures ($\mathcal{V}$, see Section \ref{sec:result} for details) and the characteristic jet angles ($\theta_\mathrm{c}$) of the $N$ GRBs. The panels correspond to the power-law (upper left), the Gaussian (upper right), and the uniform (lower left: $\theta_\mathrm{c}\equiv\theta_\mathrm{j}$; lower right: $\theta_\mathrm{c}\equiv\theta_\mathrm{j}-\theta_\mathrm{v}$) jet profile models. As references, we indicate the $p_\mathrm{3\sigma}=0.003$ (thin blue dashed lines) and the $p_\mathrm{5\sigma}=6\times10^{-7}$ (thick blue dashed lines) levels 
		in all four panels. We give the $r_{\inf}$ values the $r$ distribution medians converge to with $N$, as well as the $N\!\mathrm{s}$ where the medians ($N_\mathrm{3\sigma}$ (50\%) and $N_\mathrm{5\sigma}$ (50\%)) and the 97.8th percentiles ($N_\mathrm{3\sigma}$ (97.8\%) and $N_\mathrm{5\sigma}$ (97.8\%)) of the $p$ distributions reach $p_\mathrm{3\sigma}$ or $p_\mathrm{5\sigma}$ in Table \ref{tab:table1}.
	}
	\label{fig:hlhr}
\end{figure*}

The aim of this paper is to provide a proof-of-principle demonstration, and thus to motivate real data analysis, as well as to outline the basic steps and methods of such an analysis process in the future. 
It must be mentioned that as Figure \ref{fig:hlhr} shows, the detectability of the $\mathcal{V}-\theta_\mathrm{c}$ connection strongly depends on N, the number of GRBs with both observed gamma light curves and $\theta_\mathrm{c}$ values derived from the afterglow. It is possible that currently available datasets \citep[see e.g.][]{Wang18} do not provide a sufficient number of such GRBs for a $p<0.03$ detection of angular jet motions, even if the consequential $\mathcal{V}-\theta_\mathrm{c}$ connection exists in nature.
However also note, that if in the future, a solid connection between the gamma light curve variabilities (measured from prompt gamma light curves) and the characteristic jet angles (measured from optical afterglows) is found, this could directly provide a method for giving estimates (at least upper limits) on $\theta_\mathrm{c}$ values of GRBs solely from their prompt gamma light curves.

\onecolumn
\begin{table}
	\centering
	\caption{The $N$ sample sizes where the medians ($N_\mathrm{3\sigma} (50\%)$ and $N_\mathrm{5\sigma} (50\%)$) and the 97.8th percentiles ($N_\mathrm{3\sigma} (97.8\%)$ and $N_\mathrm{5\sigma} (97.8\%)$) of the $p$ distributions reach $p_{3\sigma}=0.003$ or $p_{5\sigma}=6\times10^{-7}$. We obtained the $r$ Pearson correlation coefficients and the corresponding $p$ values by cross-correlating the light curve variability measures ($\mathcal{V}$, see Section \ref{sec:result} for details) and the characteristic jet angles ($\theta_\mathrm{c}$) of the $N$ GRBs. We also give the $r_{\inf}$ values the $r$ distribution medians converge to with $N$. For more details on these results, see Figure \ref{fig:hlhr} and Section \ref{subseq:results}. 
	}
	\label{tab:table1}
	\begin{tabular}{lcccccr}
		\hline
		Profile & $\theta_\mathrm{c}$ & $N_\mathrm{3\sigma}(50\%)$ & $N_\mathrm{5\sigma}(50\%)$ & $N_\mathrm{3\sigma}(97.8\%)$ & $N_\mathrm{5\sigma}(97.8\%)$ & $r_\mathrm{inf}$\\
		\hline
		Power-law & $\theta_\mathrm{v}$ & $50$ & $144$ & $151$ & $286$ & $-0.40$\\
		Gaussian & $\theta_\mathrm{v}$ & $237$ & $659$ & $616$ & $>1000$ & $+0.19$\\
		Uniform & $\theta_\mathrm{j}$ & $130$ & $378$ & $327$ & $679$ & $-0.25$\\
		Uniform & $\theta_\mathrm{j}-\theta_\mathrm{v}$ & $83$ & $238$ & $148$ & $347$ &  $-0.31$\\
		\hline
	\end{tabular}
\end{table}
\twocolumn




\bibliographystyle{mnras}
\bibliography{GRBib}



%
%


\bsp	
\label{lastpage}
\end{document}